\documentclass[preprint,amsmath,amssymb,superscriptaddress,english]{revtex4-2}
\usepackage{graphicx}% Include figure files

\usepackage[linesnumbered,ruled,vlined]{algorithm2e}

\SetCommentSty{mycommfont}

\SetKwInput{KwInput}{Input}                % Set the Input
\SetKwInput{KwOutput}{Output}              % set the Output

\usepackage{graphics}% Include figure files
\usepackage{dcolumn}% Align table columns on decimal point
\usepackage{bm}% bold math professional-quality tables
\usepackage{amsfonts}       % blackboard math symbols
\usepackage{nicefrac}       % compact symbols for 1/2, etc.
\usepackage{microtype}      % microtypography
\usepackage{lipsum}
\usepackage{subcaption}
\usepackage{amsmath}
\usepackage{xcolor}
\usepackage{babel}
\usepackage{soul}
\begin{document}
\title{Certified quantum random number generator based on single-photon entanglement}

\author{Nicolò Leone \footnote{corresponding author: nicolo.leone@unitn.it}}
\affiliation{%
Nanoscience Laboratory, Department of Physics, University of Trento, Italy\\
}%

\author{Stefano Azzini}%
\affiliation{%
Nanoscience Laboratory, Department of Physics, University of Trento, Italy\\
}%

\author{Sonia Mazzucchi}%
\affiliation{%
 Department of Mathematics and TIFPA-INFN, University of Trento, Italy\\
}%

\author{Valter Moretti}
\affiliation{%
 Department of Mathematics and TIFPA-INFN, University of Trento, Italy\\
}%

\author{Lorenzo Pavesi}%
\affiliation{%
Nanoscience Laboratory, Department of Physics, University of Trento, Italy\\
}%

\begin{abstract}
Quantum entanglement represents an ideal resource to guarantee the security of random numbers employed in many scientific and cryptographic applications. However, entanglement-based certified random number generators are particularly challenging to implement. Here, we demonstrate a new certified quantum random number generator based on momentum-polarization entangled single photon states. The use of single photon entanglement allows employing an attenuated laser source and a simple setup where only linear optical components are utilized. For the latter, a semi-device-independent modeling of the photonic quantum random number generator is developed, which certifies a minimum entropy of $(2.5\pm 0.5)\%$, corresponding to a generation rate of 4.4 kHz. At the expenses of a higher level of trust in the system, the certified minimum entropy can be increased to $(30.1 \pm0.5 )\%$, implying a generation rate of 52.7 kHz. Our results show that a simple optical implementation combined with an accurate modeling provide an entanglement-based high-security quantum random number generator using imperfect devices.
\end{abstract}

\maketitle

\section{Introduction}
Random numbers represent a fundamental resource in several applications, in particular numerical simulations, internet of things and cryptography~\cite{Collantes16}. In the latter, the level of unpredictability of the sequence of random bits is a fundamental aspect, since it guarantees the security of the cryptographic protocols. For these reasons, randomness certification has become a crucial feature, i.e. being able to ensure that the random numbers are uniformly distributed, uncorrelated and unpredictable. The validation of the first two features can be done by running suitable statistical tests on the numbers sequence, while ensuring the unpredictability represents a more challenging task. In cryptography, the main figure of merit for the randomness as well as for the security of the output string is the {\em min-entropy}~\cite{Koning09}. Its certification for a random number generator (RNG) allows the use of a suitable randomness extractor to obtain from the raw bits a sequence of uniform random bits~\cite{Nisan99}.
RNGs can be classified in three main categories from the security point of view. Pseudo-random-number-generators (PRNGs) are essentially based on algorithms, hence  their security is based on assumptions on the computational power of an adversary.  True-random-number-generators (TRNGs) are based on physical phenomena which are hard to predict. Even if TRNGs are in principle more secure than PRNGs, it is rather difficult to provide a robust model of their entropy source since they are based on noisy and/or chaotic phenomena. In particular, fluctuations of the working conditions could lower significantly the entropy of the bit string. Finally, RNG based on quantum physics, where the entropy source is a quantum process, are called quantum-random-number-generator (QRNG)~\cite{Collantes16,Ma2016}.
The probabilistic nature of the measurement outcomes of quantum mechanical phenomena gives an important security advantage to QRNGs with respect to TRNGs. Moreover, this natural source of entropy can be modelled, allowing for an easy and robust estimation of the min-entropy, which is independent from the presence of additional classical noise~\cite{Frauchiger13}.
Among the different types of mechanisms for a QRNG, the {\em device-independent} QRNG is considered the most secure: the randomness certification of the generated sequence is obtained independently of any modeling of the employed devices, which are considered as black boxes. In principle, device-independent protocols should be able to handle scenarios in which device imperfections are present and adversaries can access the employed devices and program them~\cite{Acin16,Pironio18}. The main example of device-independent random number generation protocol relies on Bell inequalities violation. Indeed, it is well known that, when a Bell inequality is violated, the quantum correlation between measurements of local observables cannot be reproduced by a local hidden variable theory. In other words, the results of a quantum measurement cannot be predetermined~\cite{Bell74}, hence they are intrinsically random ~\cite{Pironio18} and unpredictable. Remarkably, it is possible to quantify the  min-entropy associated to the amount of violation of a Clauser-Horne-Shimony-Holt (CHSH) inequality~\cite{Clauser69}, thus providing an estimate of the amount of \emph{true quantum randomness} the device is able to provide~\cite{Pironio10,Acin12,Pironio13}. However, it is important to point out that the requirement of device-independence is rather demanding from an experimental point of view, requiring also loophole-free Bell tests. A few proof-of-principle experiments with entangled photon pairs have been carried out in this direction\cite{Shen18,Liu18,Bierhorst2018,Liu2021,Shalm2021}. In particular, Ref.~\cite{Shalm2021,Liu2021} have recently reported remarkable improvements in the random bit throughput, but this kind of experiment remains a technological challenge. Indeed, they typically require complicated setups having space-like separated detection stages equipped with expensive high-efficiency detectors. This makes the deployment of device-independent QRNG still prohibitive at the moment.

Throughout the years, another class of QRNG devices has been largely developed, allowing to overcome the above mentioned technological hurdles at the expense of a few assumptions on the experimental setup: these are the {\em semi-device-independent} QRNGs. In these solutions, the min-entropy is guaranteed by the fundamental principles of quantum mechanics, but they gain in ease of implementation by introducing additional assumptions on the theoretical modelling of the QRNG. This relaxes the requirements about the employed devices leading to less complicated implementations. As an example, many of them assume that one or more components of the setup are trusted, e.g. the source~\cite{cao2016,Marangon2017,Avesani2018,Thibault2019} or the measurement apparatus~\cite{Vallone2014,Cao2015,Nie2016}. Others exploit energy-constrained quantum states~\cite{himbeeck2017,rusca2019,himbeeck2019,avesani2020,rusca2020}, overlap between wavefunctions~\cite{Brask2017,Leone2020}, bounded dimensionality~\cite{li2011,lunghi2015}, quantum steering~\cite{Smith12} or quantum contextuality~\cite{deng2013,abbott2014} as means to ensure that a certain level of min-entropy is achievable. Note that even a device-independent protocol can be considered as a semi-device-independent one if just a few assumptions are introduced. The ease of implementation, the high throughput, together with the security of semi-device-independent protocols candidate this class of QRNGs to be a valuable resource for applications. 
In this paper, we report the first semi-device-independent QRNG based on single-photon entanglement (SPE), which is a particular kind of entanglement between distinct degrees of freedom of the same photon \cite{Azzini2020}. The certification scheme we demonstrate relies on the violation of a CHSH inequality using single-photon entangled states of momentum and polarization and on a model of the experimental setup. The model is based on the memory effects introduced by detectors and on the polarization dependence of the optical components (i.e. beam splitters and mirrors). The introduction of such a modeling allows to keep the experimental implementation simple. Indeed, the use of SPE presents several advantages with respect to inter-photon entanglement. Firstly, SPE states can be generated using classical light sources and off-the shelf linear optical components. Secondly, it has been reported that SPE is more robust under decoherence and dephasing \cite{Saha16}. Thirdly, due to the contextual local nature of this quantum phenomenon, neither space-like separated measurement stages nor coincidence measurements are necessary to test the CHSH inequality, allowing the use of Single Photon Avalanche Diodes (SPADs) without coincidence electronics. Therefore, our experiment represents one of the first attempts to make QRNGs based on photonic entanglement more accessible.
The paper is organized as follows: in Section \ref{Section 1}, we introduce SPE for photons and we describe the general methodology used to generate the random numbers. The experimental setup is also described. In Section \ref{Section 2}, the entropy certification protocol is analyzed focusing on the CHSH inequality. The presence of the polarization non-idealities and of the memory effects is also discussed here. In Section \ref{section 3}, the experimental data supporting our claim for a novel certified QRNG based on SPE are shown and discussed, while conclusions are finally addressed in Section \ref{section 4}, where a few future perspectives of this work are also given.
\section{Setup for quantum random numbers generation}\label{Section 1}
SPE is a type of entanglement in which a single particle, e.g. a photon, has two internal degrees of freedom entangled. We consider single photon entangled states of the form:
\begin{equation}\label{eq:SPE-state}
|\psi\rangle=\frac{1}{\sqrt{2}}\left(|0V\rangle +|1H\rangle\right),   
\end{equation}
where the photon momentum ( $|0\rangle$ and $|1\rangle$ ) is correlated with the photon polarization ( $|V\rangle$ and $|H\rangle$ ). The wavefunction $|\psi\rangle$ belongs to the space $\mathbb{C}^2_M \otimes \mathbb{C}^2_P$, with obvious meaning of the subscripts.

\begin{figure}[h!]
\centering
\includegraphics{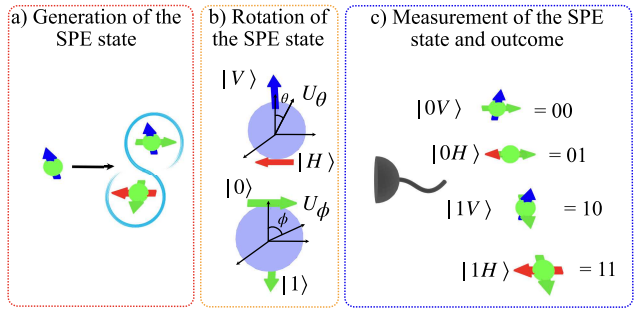}
\caption{Method to generate quantum random numbers: a) a single photon entangled (SPE) state is generated; b) the two degrees of freedom of the SPE state are rotated by $\phi$ and $\theta$, through unitary rotation operators ($U_{\phi},U_{\theta}$), represented on the Bloch sphere; c) Single photon avalanche diodes (SPADs) measure the rotated SPE state, i.e. the result of the SPE state projection on one of the four basis states ($|0V\rangle,|1V\rangle,|0H\rangle,|1H\rangle$); two bits are generated according to a coding which identifies the SPAD which has measured the photon. Therefore, a sequence of random numbers is the outcome of a sequence of quantum state projections.}
\label{fig:Methods}
\end{figure}
The generation mechanism of the random numbers is schematized in Figure~\ref{fig:Methods}. First, a SPE state of single photon of the form (\ref{eq:SPE-state}) is generated (Figure~\ref{fig:Methods}a). Second, the state undergoes separate rotations in momentum and polarization by angles $\phi$ and $\theta$, respectively (Figure~\ref{fig:Methods}b). Third, the resulting state is measured and a random symbol is generated, according to which detector clicks (Figure~\ref{fig:Methods}c). The randomness is intrinsic to the quantum measurement process: the rotations modify the expectation values of the projector operators over the four different states $|0V\rangle,|1V\rangle,|0H\rangle,|1H\rangle$ composing the Hilbert space $\mathbb{C}^2_M \otimes \mathbb{C}^2_P$. Then, by repeating the procedure $n$ times, each time by varying the couple of angles ($\phi$, $\theta$), a sequence of $n$ random symbols is obtained. Finally, the symbols are translated into binary numbers according to a coding (Figure~\ref{fig:Methods}c). This random number generation mechanism is the outcome of the projection operation of the state (\ref{eq:SPE-state}) over a rotated basis corresponding to a couple $(\phi$, $\theta)$. 

\begin{figure}[h!]
\centering
\includegraphics{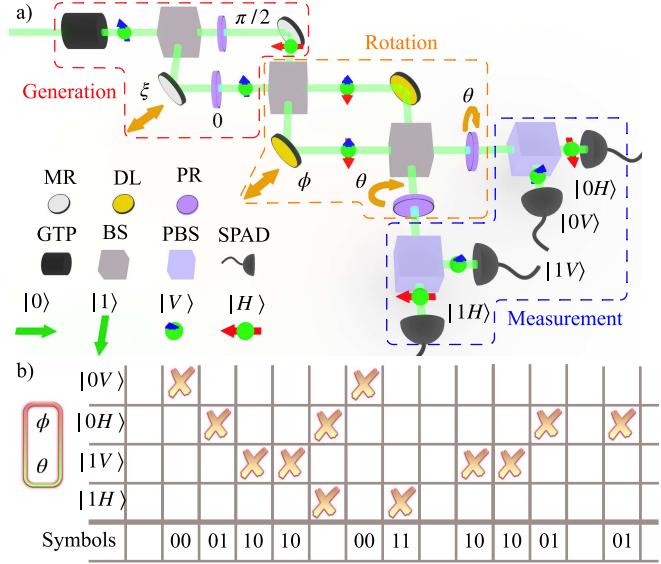}

\caption{a) Experimental setup used to generate random numbers. It is composed by the generation (red box), the rotation (orange box) and the detection (blue box) stages. The green line shows the optical path of the photons, represented by a tiny green sphere. The photon polarization degree of freedom is indicated by vertical blue and horizontal red arrows. Green arrows represent the photon momentum. The used linear optical components are labeled according to:  GTP, Glan-Thompson polarizer; BS, beam splitter; MR, mirror; DL, delay line (three mirrors, see Supplementary Note 2); PBS, polarized beam splitter; PR, polarization rotator (half-wave plate); SPAD, single photon avalanche diode. $\xi$: angle used to correct for phase differences in the generation. $\phi$: momentum rotation angle. $\theta$: polarization rotation angle. b) Example of a generated raw random number sequence. This is produced by translating in bits the temporal sequence of symbols generated by coding the event "which SPAD detected the photon", given a certain couple ($\phi$,$\theta$). Multiple detection events and time bins with no detection are discarded.}
\label{fig:Setup}
\end{figure}

To implement this, we use the setup of Figure~\ref{fig:Setup}a. The state $|\psi\rangle$ is generated by using linear optical components and by operating on the single photons emitted by an attenuated green He-Ne laser  \cite{Pasini20}. First, a Glan-Thompson polarizer fixes the initial polarization state to $|0V\rangle$, namely single photons propagating through the $|0\rangle$ direction with $|V\rangle$ polarization. Then, by using a beam splitter, two half-wave plates and two mirrors, the desired SPE state (\ref{eq:SPE-state}) is formed (red box in Figure~\ref{fig:Setup}a). Two waveplates are used to get equal phase retardation, one is rotated at $\pi/2$ angle to rotate the polarization, the second is placed at $0$ angle. Additional phase mismatches are compensated adjusting the phase $\xi$ by moving one of the mirror. Note that, by exchanging the role of the wave plates in the paths and/or by setting the proper phase $\xi$, it is possible to obtain any Bell state.

The rotation of the SPE state is obtained by a Mach-Zehnder interferometer (MZI) and two half-wave plates (orange box in Figure~\ref{fig:Setup}a).
The MZI rotates the momentum by an angle $\phi$, while the two half-wave plates rotate the polarization by an angle $\theta$. Note that the actual phase difference between the two arms of the MZI is 2$\phi$.

The measurement of the rotated SPE state is done by using the two output ports of the MZI, two polarized beam splitters and four SPADs (blue box in Figure~\ref{fig:Setup}a), whose properties are well known (trusted detectors in terms of efficiency, dark counts, dead time and afterpulsing). The measure projects the rotated SPE state on the four possible single photon states $|0V\rangle,|0H\rangle,|1V\rangle,|1H\rangle$. The sequence of single photon detection events is registered by a time-tagging electronics and corresponds to the sequence of random numbers produced by our setup (Figure~\ref{fig:Setup}b). 

By using the same setup, Bell's inequality violation has witnessed entanglement between the momentum and the polarization degrees of freedom of the SPE states (\ref{eq:SPE-state}) \cite{Pasini20}. It is here relevant to observe that, in the case of single-particle entanglement, the intrinsic randomness is due to contextuality, and not to non-locality, as it is the case of inter-particle entanglement between two distinct particles \cite{Azzini2020}. 

\section{Protocol for entropy certification}\label{Section 2}
Having generated a sequence of numbers by a quantum process, the next step is to certify their randomness. This is usually done by evaluating the amount of quantum entropy intrinsic to the generated sequence \cite{Collantes16}. A protocol to estimate the amount of quantum randomness in a device-independent QRNG exploiting the quantum correlations of a non-local entangled state between two space-like separated particles was proposed for the first time in \cite{Pironio10}. It is based on two different local measurement operations ($U_x^1,U_y^2$), applied to each particle, that give as output two binary results called $a$ and $b$. These two operations depend on the value of two input bits $(x,y)$, provided by another QRNG.
Accumulated a sequence $R=(a_1,b_1;\ldots;a_n,b_n)$ of $n$ outcomes of the form $(a,b)$ for a certain input random sequence $S=(x_1,y_1;\ldots; x_n, y_n)$ of bits $(x,y)$, the quantum correlations are evaluated by the CHSH correlation function:
\begin{equation}\label{eq:CHSHfun}
    I=\sum_{x,y}(-1)^{xy}(\mathbb{P}(a=b|x,y)-\mathbb{P}(a\neq b|x,y)),
\end{equation}
where $\mathbb{P}(a=b|x,y)$ is the probability to measure $a=b$ given $(x,y)$ and $\mathbb{P}(a\neq b|x,y)$ is the probability to measure $a\neq b$ given $(x,y)$. Systems with a classical deterministic description satisfy $|I|\leq2$, while certain measurements performed on entangled states violate this inequality, and the randomness of the sequence $R$ is ensured. Moreover, the min-entropy $H_{min}(R|S)$ of the outcome sequence $R$, given the initial sequence $S$, is given by \cite{Pironio10}:
\begin{equation}\label{eq:H_min}
     H_{min}(R|S) = -n\log_2 \left[\mathbb{P}_{guess}(a,b|x,y)\right],
\end{equation}
where the guessing probability $\mathbb{P}_{guess}(a,b|x,y)$ is the largest probability to measure any outcome $(a,b)$ given any $(x,y)$. This is upper-bounded by \cite{Pironio10}
\begin{equation}\label{eq:p_guess}
     \mathbb{P}_{guess}(a,b|x,y)\leq \frac{1}{2}+\frac{1}{2}\sqrt{2-\frac{I^2}{4}}.
\end{equation}
The inequality \eqref{eq:p_guess} gives an estimation of the effectiveness of the optimal strategy to guess the sequence $R$ knowing the sequence $S$, while  \eqref{eq:H_min} provides an upper bound for the number of uniform random bits that can be extracted from the raw sequence of data $R$ \cite{Koning09}. Randomness extraction is done by using known techniques \cite{Nisan99}, such as the Toeplitz extractor \cite{Ma13}.

In our experiments, the angles $(\phi,\theta)$ play the role of the inputs $(x,y)$, therefore we labeled them $(\phi_x,\theta_y)$ when we set them to evaluate \eqref{eq:CHSHfun}. Then, for each measurement choice $(\phi_x,\theta_y)$, a couple of outcomes $(a,b)$ is produced, corresponding to a detection event in the path $a$ with polarization $b$, where we identify $a=0$ to $|0\rangle$, $a=1$ to $|1\rangle$, $b=0$ to $|V\rangle$ and $b=1$ to $|H\rangle$. Measured a sequence of outcomes, i.e. a sequence of single photon detection events for a given input sequence $(\phi_x,\theta_y)$, the Bell violation is evaluated through an estimator $\hat{I}$ of \eqref{eq:CHSHfun}:
\begin{equation}\label{eq:parCHSH}
\begin{aligned}
    &\hat{I}(\phi_0,\phi_1,\theta_0,\theta_1)=\\
    &=\sum_{x,y}(-1)^{xy}(\mathbb{\hat{P}}(a=b|\phi_x,\theta_y)-\mathbb{\hat{P}}(a\neq b|\phi_x,\theta_y)),
\end{aligned}
\end{equation}
the probabilities $\mathbb{\hat{P}}(a,b|\phi_x,\theta_y)$ are computed as the maximum likelihood estimators of the probabilities $\mathbb{P}(a,b|\phi_x,\theta_y)$.
Since the inputs $(\phi_x,\theta_y)$ are set during a scan, memory effects could correlate successive measurements outcomes. Memory might be caused by the non-idealities of the SPADs, e.g. dead time and afterpulsing, and by the use of an attenuated light source. These yield stochasticity of the photons arrival times. Because of this, $\mathbb{\hat{P}}(a,b|\phi_x,\theta_y)$ is estimated by a Markovian model parameterized on the average photon flux detected by the SPADs, the afterpulsing probability and the SPAD dead time (see Supplementary Note 4 and \cite{Mazzucchi2021}). The resulting guessing probability is defined $\mathbb{P}_{guess}^{*}(a,b|\phi_x,\theta_y)$. In our setting, SPAD dark counts are negligible compared to the photon flux from the source.

The  protocol described in \cite{Pironio10} relies on a loophole-free Bell test, hence it requires to randomly switch the observables to be measured. This requirement is important in a device-independent scenario where an eavesdropper is allowed to control the detectors. Indeed, in this case the  knowledge of the measurement basis would allow  Eve to program the devices in a way that would mimic Bell inequality violation. However, if we assume that the detectors provider is trusted, the semi-device-independent protocol is essentially aimed to provide an estimate of min-entropy robust against unwanted flaws of the system. In this regard, the random switch of measurement basis is not necessary, since Eq. \eqref{eq:p_guess} is robust under classical side information, the latter including also the choice of the measurement basis. In addition, in the device independent protocol presented in ref.\cite{Pironio10}, the random choice of  $\phi$ and $\theta$ plays an important role in the construction of an estimator for the Bell parameter $S$, allowing to tackle memory effects in the experimental devices. In the present work, these issues are addressed  by constructing estimators for the 16 quantum probabilities present in $S$ by means of the Markov model described in the Supplementary Note 4.

The last issue to compute \eqref{eq:parCHSH} by using the SPE state as in \eqref{eq:SPE-state} is the possible existence of a communication channel between the two degrees of freedom. In the setup, non-idealities of the beam splitters and of the mirrors of the MZI make momentum and polarization degrees of freedom no longer independent. Therefore, the actual characteristics of the optical elements (polarization-dependent reflectance, transmittance and absorption) have to be considered. 
Specifically, projection-valued measures describing the measurement operation cannot be written in the product form $\{P_{\phi_x}\otimes P_{\theta_y}\}_{x,y=0,1}$, where $P_{\phi_x}$ and $P_{\theta_y}$ are the projection operators for a given $(\phi_x,\theta_y)$.
To deal with this, we numerically evaluate an upper bound $e_P$ for the difference between the ideal probabilities obtained by $\{P_{\phi_x}\otimes P_{\theta_y}\}$ and the estimated probabilities (named real probabilities in the following) obtained by modeling the experimental setup (see Supplementary Note 3).
$e_P$ is calculated by considering any possible input state $\rho \in \mathbb{C}^2_M \otimes \mathbb{C}^2_P$, with $\phi,\theta\in [0,2\pi]$. As a consequence, an upper bound $e_I$ to the difference $|\hat{I}_{\text{ideal}}- \hat{I}_{\text{real}}|$ between the ideal and the real CHSH correlation functions is computed as well. 
Thus, as long as these bounds are satisfied, \eqref{eq:p_guess} becomes
\begin{equation}
	\mathbb{P}_{guess}(a,b|\phi_x,\theta_y) \leq  \frac{1}{2}+\frac{1}{2}\sqrt{2-(|\hat{I}_{real}|-e_I)^2/4} + e_P .
\end{equation}
Eventually, taking into account the memory effects, $\mathbb{P}_{guess}^{*}(a,b|\phi_x,\theta_y)$ is estimated as 
\begin{equation}\label{eq:corrected_H}
\begin{aligned}
	&\mathbb{P}_{guess}^{*}(a,b|\phi_x,\theta_y) \leq\\
	&\leq\mathbb{M}\left( \frac{1}{2}+\frac{1}{2}\sqrt{2-(|\hat{I}_{real}|-e_I)^2/4} + e_P\right), 
\end{aligned}
\end{equation}
where $\mathbb{M}$ is a function which results from the Markovian model (see Eq. (S30) in Supplementary Note 4) \cite{Mazzucchi2021}.
 From $\mathbb{P}_{guess}^{*}$, we calculate $H_{min}^*$ associated to each measurement outcome as
\begin{equation}\label{eq:Hmin_exp-0}
    H_{min}^* = -\log_2\left[\mathbb{P}^{*}_{guess}(a,b|\phi_x,\theta_y)\right] ,
\end{equation}
and the min-entropy of the whole sequence $R$ as
\begin{equation}\label{eq:Hmin_exp}
\begin{aligned}
    H_{min}(R|S) &=n H_{min}^*\\ &=-n\log_2\left[\mathbb{P}^{*}_{guess}(a,b|\phi_x,\theta_y)\right].
    \end{aligned}
\end{equation}
As far as $H_{min}(R|S)>0$, we can extract an unbiased sequence of random numbers from the raw data.

\begin{figure}[h!]
	\centering
	\includegraphics[scale=0.8]{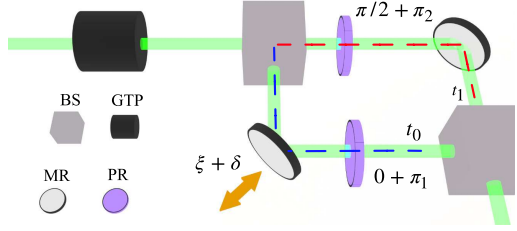}
	\caption{Generation stage of the setup with indicated the non-idealities of the optical components. $\delta$ additional phase-shift between the two paths. $\pi_1$ and $\pi_2$ additional angles of rotation of the waveplates. $t_0$ and $t_1$ transmission coefficients of the blue dashed and red dashed paths, respectively. The same optical component symbols as in Figure~\ref{fig:Setup} have been used.}
	\label{fig:eve}
\end{figure}

\begin{figure}[h!]
	\centering
	\includegraphics{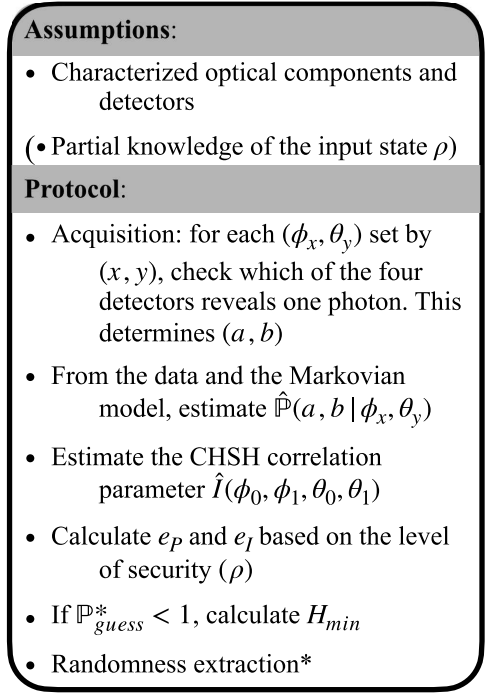}
	\caption{Schematic summary of the assumptions and the main steps of the protocol for calculating the min entropy. (*Please note that despite this last step being necessary for extracting the sequence of random numbers, it is not implemented in our experiment).}
	\label{fig:Protocols}
\end{figure}

Furthermore, it is possible to increase $H_{min}^*$ by introducing a few assumptions on the unknown input state $\rho$ which lower the two upper bounds $e_I$ and $e_P$. Let us assume to have non-ideal optical components (see Figure \ref{fig:eve} ) and the presence of an eavesdropper. In this case, at the input of the first beam splitter of the MZI (orange box in Figure~\ref{fig:Setup}a), $\rho$ can be modeled as:
\begin{equation}\label{eq:Model_Rho}
\rho(v,\delta,\pi_1,\pi_2)=R(\pi_1,\pi_2)\rho_s(v,\delta)R(\pi_1,\pi_2)^{\dagger}.
\end{equation}
$R(\pi_1,\pi_2)$ represents unwanted rotation of the wave plates by unknown angles $\pi_1\in[0,2\pi]$ and $\pi_2\in[0,2\pi]$ in the generation stage (red box of Figure~\ref{fig:Setup}a).
$\rho_s(v,\delta)$ is the actual entangled state:
\begin{equation}
\rho_s(v,\delta)=v\left(|\psi(\delta)\rangle\langle\psi(\delta)|\right)+\frac{1-v}{4}\mathbb{I}_{4},
\end{equation}
where $v\in[0,1]$ is a visibility parameter which accounts for the non-ideality of the setup and the different sources of noise \cite{Pasini20}, $\delta\in[0,2\pi]$ an additional phase,
\begin{equation}
|\psi(\delta)\rangle=\left(t_{0n}|0V\rangle + t_{1n}e^{i\delta}|1H\rangle\right),
\end{equation}
\begin{equation}\label{eq:ncoeff}
    t_{0n}=\frac{t_{0}}{\sqrt{t_{0}^2+t_{1}^2}}, \qquad 
    t_{1n}=\frac{t_{1}}{\sqrt{t_{0}^2+t_{1}^2}},
\end{equation} 
and $\mathbb{I}_{4}$ is the identity matrix. The $\pi_1,\pi_2,\delta, v$ values could be due to the characteristics of the optical components or to the action of the eavesdropper.
$t_{0}$ and $t_{1}$ are the transmission coefficients of the optical paths $|0\rangle$ and $|1\rangle$ as shown in Figure~\ref{fig:eve}. $t_{0n}$ and $t_{1n}$ are the normalized transmission coefficients which account for the probability of photon transmission on a path with respect to the probabilities that the photon is transmitted on anyone of the two paths.
Every time one of the ${t_{0n},t_{1n},v,\delta,\pi_1,\pi_2}$ parameters is fixed, a new assumption on the QRNG is introduced reducing its generality. Indeed, 
the eavesdropper could design an attack to change $\rho$ with respect to \eqref{eq:Model_Rho}, potentially jeopardizing the overall security. However, as long as an attack of this type is not feasible or is highly unlikely, the knowledge of the values of these parameters can increase $H_{min}^*$ with uncompromising security. In the ideal case, neither experimental errors nor eavesdroppers are present, and the state (\ref{eq:Model_Rho}) reduces to
\begin{equation}\label{eq:ideal_rho}
    \rho(v=1,\delta=0,\pi_1=0,\pi_2=0)=|\psi(0)\rangle\langle\psi(0)|.
\end{equation}
Finally, our entropy certification protocol, being the result of a modeling based on the characteristics of the measurement setup \cite{Mazzucchi2021}, can be classified as semi-device-independent, providing a fair lower bound to the amount of measurable $H_{min}$. The protocol and the relative assumptions are summarized in Figure \ref{fig:Protocols}.

\begin{figure}[h!]
	\centering
	\includegraphics[scale=0.6]{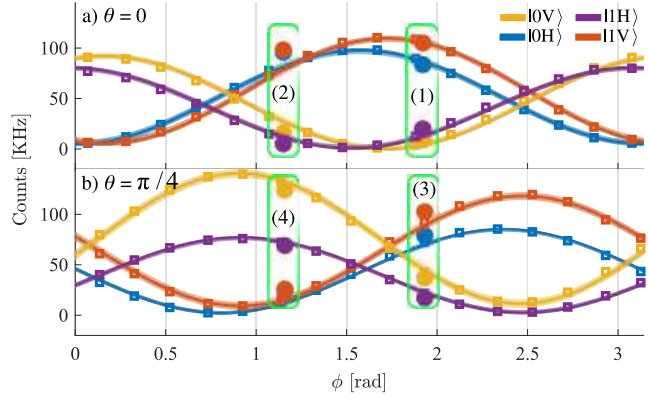}
	\caption{
		Measured count rates (empty squares) as a function of $\phi$ and for a) $\theta=0$, and b) $\theta=\frac{\pi}{4}$. The colors refer to the different states: yellow $|0V\rangle$, purple $|1H\rangle$, blue $|0V\rangle$ and red $|1V\rangle$. The solid lines are a sinusoidal fits (the $99\%-$confidence intervals is shown by the shaded area around the solid line). The green boxes highlight the working points at which the 50 s long data acquisitions are performed. (1) corresponds to $(\phi_1,\theta_0)$, (2) to $(\phi_0,\theta_0)$, (3) to $(\phi_1,\theta_1)$ and (4) to $(\phi_0,\theta_1)$. The measured total count rates at these points are represented by the solid dots.}
	\label{fig:CountsR}
\end{figure}

\section{Experimental demonstration of a certified QRNG}\label{section 3}

\begin{figure}[h!]
\centering
\includegraphics{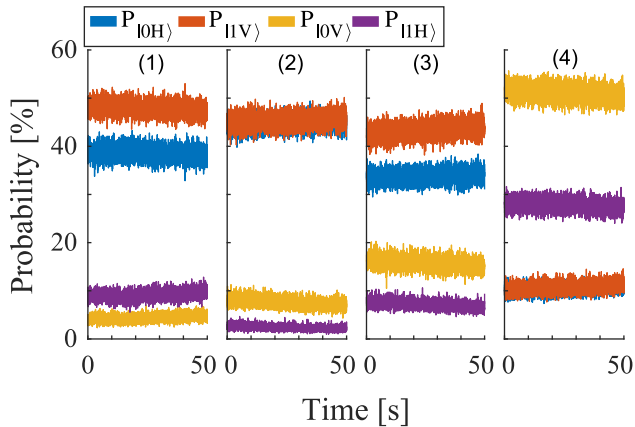}
\caption{Probabilities as a function of time for each measurement outcome (yellow $|0V\rangle$, purple $|1H\rangle$, blue $|0V\rangle$ and red $|1V\rangle$) at the four working points (1), (2), (3), (4) of Figure \ref{fig:CountsR} and Table \ref{tab:counts}, corresponding to $(\phi_1,\theta_0)$, $(\phi_0,\theta_0)$, $(\phi_1,\theta_1)$, $(\phi_0,\theta_1)$, respectively. The estimates have been done considering time intervals of 10 ms.}
\label{fig:Prob}
\end{figure}

The maximum violation of the Bell inequality is obtained for $\{(\phi_x,\theta_y)\}_{x,y=0,1}$, where $\phi_0=\frac{3}{8}\pi$ and $\phi_1=\frac{5}{8}\pi$ for momentum, $\theta_0=0$ and $\theta_1=\frac{\pi}{4}$ for polarization are chosen \cite{Pasini20}. To set these values, we fixed $\theta_{0,1}$ and vary $\phi$ by a piezoelectric transducer actuated mirror (Figure~\ref{fig:Setup}), so that sinusoidal sequences of counts (empty squares in Figure~\ref{fig:CountsR}) are measured and the $\phi$ calibration is obtained  (solid lines in Figure~\ref{fig:CountsR}). Then, four time sequences of single photon count rates for each couple $(\phi_x,\theta_y)$ are acquired for 50 s with time-bins of $1$ $\mu$s (solid dots inside green boxes in Figure~\ref{fig:CountsR}). From these sequences, the clicking SPAD is identified in each time-bin and the corresponding symbol is stored in the random symbol time sequence (Figure \ref{fig:Setup}b). Multi-photon detection events within the same time-bin are removed: they constitute the $(12.0 \pm 0.4)\%$ of the raw data and are mainly due to the statistics of emission of the source.
\begin{table}[h]
\centering
\begin{tabular}{c|cccc}
\hline
Channel &
$(\phi_0,\theta_0)(2)$ & 
$(\phi_1,\theta_0)(1)$ &
$(\phi_0,\theta_1)(4)$ & 
$(\phi_1,\theta_1)(3)$ \\
\hline
$|0V\rangle$ &  643132  &   371255    & 4754594 & 1426837   \\
$|1H\rangle$ &  202823  &   779771    & 2589956 & 652294    \\
$|0H\rangle$ &  3804170 &   3311003   & 964121  & 3078159   \\
$|1V\rangle$ &  3855004 &   4108774   & 996276  & 3945250   \\
\hline
Total &  8505129 &   8570803   & 9304947  & 9102540   \\
\end{tabular}
\caption{\textbf{Experimental counts.} The number of counts acquired during the 50 s long acquisitions for the four experimental realizations $\{(\phi_x,\theta_y)\}_{x,y=0,1}$. The numbers ${(i)=1,2,3,4}$ refer to the different green boxes in Figure~\ref{fig:CountsR}.}
  \label{tab:counts}
\end{table}
The whole numbers of counts per each SPAD acquired over 50 s integration windows are reported in Table~\ref{tab:counts}. The relative raw probabilities, estimated over a 10 ms acquisition time, are plotted in Figure~\ref{fig:Prob} for each $(\phi_x,\theta_y)$. The observed slight probability increase or decrease can be attributed to the temporal stability of the setup, e.g. the piezoelectric transducer actuated mirrors. This affects $H_{min}$ but it is mitigated by the certification protocol. More stable setup could result in a sequence of random numbers with higher $H_{min}$. 
Table~\ref{tab:prob} shows the average raw probabilities. These have been estimated by dividing the 50 s acquisition interval in 10 s sub-intervals, computing the probabilities in the sub-intervals and getting the mean of these. This procedure yields also a $\simeq$ 0.2 $\%$ standard error on the raw probabilities. Table~\ref{tab:prob} also reports within parentheses the average probabilities $\mathbb{\hat{P}}(a,b|\phi_x,\theta_y)$ corrected by the Markovian model, which are equal to the raw probabilities within the errors. $\mathbb{\hat{P}}(a,b|\phi_x,\theta_y)$ are used to compute $\hat{I}$, for which we obtain:
\begin{equation}\label{eq:CHSH_exp}
    |\hat{I}(\phi_0,\phi_1,\theta_0,\theta_1)|=2.656\pm0.003 .
\end{equation}
From $\hat{I}(\phi_0,\phi_1,\theta_0,\theta_1)$ and considering an ideal setup, we get an overall certified min-entropy $ H_{min}(R|S)=42.8\pm 0.4\%$, \eqref{eq:Hmin_exp}.  

\begin{table}[h!]
\centering
\begin{tabular}{c|cccc}
\hline
Channel &
$(\phi_0,\theta_0)(2)$ & 
$(\phi_1,\theta_0)(1)$ &
$(\phi_0,\theta_1)(4)$ & 
$(\phi_1,\theta_1)(3)$ \\
\hline
$|0V\rangle$ &  7.6(7.6) &   4.3(4.3)   & 51.1(51.1)  & 15.7(15.7)   \\
$|1H\rangle$ &  2.4(2.4) &   9.1(9.1)   & 27.8(27.8)  & 7.2(7.2)   \\
$|0H\rangle$ &  44.7(44.7) &   38.6(38.6)   & 10.4(10.4)  & 33.8(33.8)   \\
$|1V\rangle$ &  45.3(45.3) &   47.9(48.0)   & 10.7(10.7)  & 43.3(43.4)   \\
\hline
\end{tabular}
\caption{\textbf{Measured raw and corrected probabilities.} Mean values of the experimental $({\tilde{\mathbb{P}}})$ probabilities for the four couples of angles $\{(\phi_x,\theta_y)\}_{x,y=0,1}$. Within parenthesis the maximum-likehood $({\hat{\mathbb{P}}})$ probabilities are reported which result from the Markovian model. Errors on the probabilities of $\simeq 0.2\%$ are estimated based on repeated measurements. The numbers within parenthesis in the columns header refer to the  measurement points shown in Figure \ref{fig:CountsR}.}
  \label{tab:prob}
\end{table}
\begin{table*}
\centering
\caption*{}
\begin{tabular}{cc|cccc}
\hline
Variable & Fixed&
$e_P\cdot 10^{-2}$ & 
$e_I\cdot 10^{-2}$ &
$H^*_{min}$ &
Random bits generation rate [kHz]
\\
\hline
$\rho$ general & - & $8.0\pm0.2$  &  $33.2\pm0.8$   &   $(2.5\pm 0.5)\%$ & 4.4  \\
$\delta,\pi_1,\pi_2$, $v$ &$t_{0n},t_{1n}$& $8.0\pm0.2$ & $26.4\pm0.8$ &  $(6.3\pm 0.6)\%$ & 11.0  \\
$\delta$, $v$&$t_{0n},t_{1n},\pi_1,\pi_2$  & $7.8\pm0.2$ & $1.2\pm 0.2$ & $(26.9\pm 0.5)\%$ & 47.1   \\
 $v$ &$t_{0n},t_{1n},\pi_1,\pi_2,\delta$&$6.6\pm0.2$ & $0.26\pm 0.07$ &  $(30.1\pm 0.5)\%$ &52.7  \\
\hline
\end{tabular}
\caption{\textbf{Min-entropy $H^*_{min}$ and random bits throughput for different levels of trust of our SPE-based QRNG.} In the table is reported the value of $e_P,e_I$, the min-entropy $H^*_{min}$ and the random bit throughput (assumed instantaneous extraction procedure) for different level of model for the input state $\rho$. Each row corresponds to a level of modelling, starting from the most general. In the first column there are the parameters over which the maximization of $e_P$ and $e_I$ is performed. In the second column, we reported the parameters fixed the model. In the second row $t_{0n},t_{1n}$ are fixed following \eqref{eq:ncoeff} and using the experimental value measured. In the third row, we fix $\pi_1=\pi_2=0$. Lastly, in the fourth row where $\delta$ is fixed to 0.}
  \label{tab:h_min}
\end{table*} 

However, if we consider the setup non-idealities, we find $e_I=0.332 \pm 0.008$ and $e_P=0.080 \pm 0.002$ which yield a  $H_{min}^*=(2.5 \pm 0.5)\%$, according to \eqref{eq:corrected_H} and \eqref{eq:Hmin_exp-0}. This number results from the assumption of a completely untrusted or unknown input state and, therefore, represents the most conservative, i.e. most secure, estimate. Note that the upper bound to $\mathbb{P}_{guess}(a,b|x,y)$ given by \eqref{eq:p_guess} represents the best estimate for the marginal guessing probability, e.g. $\mathbb{P}_{guess}(b|y)=\max\sum_a \mathbb{P}_{guess}(a,b|x,y)$, where the marginal refers to a single degree of freedom. Therefore, from an operative point of view, we write as bit $b=0$ each photon detected with vertical polarization $|V\rangle$ independently of its momentum state, and as bit $b=1$ each photon detected with horizontal polarization $|H\rangle$ independently of its momentum state. This means that, having acquired a total amount of $\simeq35\times 10^6$ raw data, we can obtain $\simeq 0.88\times 10^6$ certified random bits after the extraction procedure. Considering that the entire acquisition process requires 200 s, we can get a certified random bits generation rate $\simeq 4.4$ kHz, neglecting the randomness extraction phase.

Table~\ref{tab:h_min} shows the increase in $H_{min}^*$ as we increase the level of trust on the input state. Four different levels of trust are considered and the resulting upper bounds $e_I$ and $e_P$ together with the corresponding minimum entropies $H_{min}^*$ and random number generation rates are reported. Fixing more parameters of the input state \eqref{eq:Model_Rho} means a higher knowledge of the working conditions of the setup, i.e. a better confidence level or trust, which means making more assumptions on the overall behavior of the system. 
In the fully trusted setup, i.e. when we know all the parameters but the visibility $v$, \eqref{eq:Model_Rho} becomes
\begin{equation}
\begin{aligned}
    &\rho(v,\delta=0,\theta_1=0,\theta_2=0)=\\
    &=v\left(|\psi(0)\rangle\langle\psi(0)|\right)+\frac{1-v}{4} \mathbb{I}_{4},
\end{aligned}
\end{equation}
which differs from the ideal case, \eqref{eq:ideal_rho}, for the presence of a parameter $v\leq 1$. In this fully trusted case, the maximum amount of min-entropy $H_{min}^*=(30.1\pm0.5)\%$. 

If the control of the phase $\delta$ is removed, $\delta$ is free to vary within $[0,2\pi]$. This could be considered as an additional phase coming from an unwanted erroneous or fraudulent control of the phase $\xi$ (see Figure~\ref{fig:Setup}). In this case, we obtain $H_{min}^*=(26.9\pm 0.5)\%$. Then, we can relax also the assumptions on the two polarization rotators (wave plates) present in the generation stage: an external eavesdropper could selectively rotate each one of them by an angle $\pi_1$ and/or $\pi_2$. This leads to $H_{min}^*=(6.3\pm 0.6)\%$. 

\section{Conclusion} \label{section 4}
In this work, we present a certified quantum random number generator (QRNG) exploiting single-photon entangled states from an attenuated laser beam. Our results show that a simple setup combined with an accurate modeling of its optical elements and detectors can provide an entanglement-based high-security QRNG using imperfect devices. 
Our protocol can be classified as  semi-device independent since the entropy certification scheme relies on a certain number of assumptions. In particular, the model relies on the detailed knowledge of the technical specifications of the optical components of the setup and of the detectors. The first ones are necessary to estimate $e_P$ and $e_I$, the second one are related to some parameters present in the Markovian model. No additional assumptions on the state of the photons entering in the setup and on the detailed behaviour of the measuring apparatus are needed, yielding an estimate of min-entropy robust under unwanted flaws of the system. We remark that, due to the less-than-unity detectors efficiency, the model implicitly relies on the fair sampling assumption. In addition, we assume that preparation and measurement parameters of the system are stable during the acquisition time.

We demonstrate a certified quantum minimum entropy $H_{min}^*=2.5\%$, whose value can be increased to $30.1\%$ by increasing the level of trust in the experimental setup. 
The certification relies on the violation of a Bell inequality in the CHSH form. In our certification scheme, we take into account the non-idealites of our measurment setup: the presence of afterpulsing and dead time of SPADs, together with the polarization non-idealities of beam splitters and mirrors composing the Mach-Zenhder interferometer. The detectors introduce, in the sequence of outcomes, memory effects which are considered by a Markovian model, finally resulting in negligible corrections to the estimated probabilities. The polarization-based non-idealities are considered by calculating two upper bounds, $e_I$ and $e_P$, of the difference between the ideal correlation function/probabilities and the measured ones. This accurate modeling allows one considering either a most secure scenario (less trusted) where minimum assumptions are done, or a trusted setup scenario where the full knowledge of the components and of their setting is assumed. Consequently, we can move from a certified random number generation rate of 4.4 kHz to 52.7 kHz. 

Despite these rates do not represent an improvement compared to other much faster semi-device independent QRNGs \cite{Brask2017,rusca2019,rusca2020}, we stress that the novelty of our work is that of being the first experimental demonstration of a practical QRNG certifying the min-entropy based on single photon entanglement and violation of a Bell inequality. Nevertheless, the rate can be further increased by enhancing the laser photon flux up to the linearity limit of the SPADs, i.e. 1 MHz, and/or by reducing the time bin duration in the acquisition, which decreases the number of multi-photon detection events rejected by the protocol. With these improvements, we estimate a random bits generation of up to 500 kHz for the fully trusted case with the actual experimental setup. Having access to detectors with a lower dead time will surely improve even more the rate of the QRNG. Furthermore, the use of degrees of freedom in a setup without possible communication channels or the use of higher quality optical components will lower the upper bounds $e_I$ and $e_P$, resulting in a higher value of min-entropy up to the ideal value of almost $43\%$. In particular, the improvements highlighted above can find a natural implementation in a suitably designed integrated photonic platform, whose compactness and ease of integration with electronics will help making our novel type of photonic QRNG a potentially high-speed deployed device.

\section*{Methods}
The measurements are performed using as a light source an attenuated single mode green He:Ne laser, emitting at 543.5 nm, with nominal output power of 4 mW. The laser is fiber-coupled and attenuated by a variable optical attenuator before entering the experimental setup. The phase $\xi$ and $\phi$ are controlled by using two piezoelectric transducer actuated mirrors with feedback loop to ensure time stability. Four lenses collect the four projection-valued measures and fiber-couple them to four different single photon avalanche diodes (SPADs), whose efficiencies have been previously equalized using four variable optical attenuators, which compensate also for different losses due to fibers and lenses. For each experimental realization $\{(\phi_x,\theta_y)\}_{x,y=0,1}$, the observation windows of 50 s are measured using time bins of $1\mu$s. Time bins with multiple detection events or no detection at all are discarded from the collected string of raw events. Except for the laser, the whole setup is optically shielded from the rest of the lab by enclosure in a black box. In addition, each SPAD is optically shielded and physically separated from the three others by means of opaque boxes, in order to avoid any cross-talk in the measurements.

\section*{Funding}
Project H2020 820405 QRANGE. Q@TN from PAT(AI).
\section*{Acknowledgments}
We acknowledge helpful discussions with C. Agostinelli on the code implementation of the Markovian model. This project has received funding from the European Union’s Horizon 2020 research and innovation
programme under grant agreement No 820405. NL was supported by a fellowship of Q@TN within the PAT(AI) grant.
\section*{Author contributions}
The original experiment was conceived by all the authors. N.L. and S.A. performed the experiment in the laboratory, under the supervision of L.P. S.M. and V.M. developed the Markovian model used in the experiment. S.M. and N.L. carried out the analysis of the imperfection with the help of V.M and S.A. All the authors discussed the results and contributed to the writing of the manuscript.
\section*{Disclosures} L.P., V.M., S.M. declare the following competing interests: a patent has been filed on the SPE.

\section*{Data availability} Data underlying the results presented in this paper are not publicly available at this time but may be obtained from the authors upon reasonable request.

\section*{Supplemental Information}
See Supplementary Information for supporting content. 

% Bibliography
\bibliography{Reference}

\end{document}

\title{Supplementary Information: Certified quantum random number generator based on single-photon entanglement}

\author{Nicolò Leone \footnote{corresponding author: nicolo.leone@unitn.it}}
\affiliation{%
Nanoscience Laboratory, Department of Physics, University of Trento, Italy\\
}%

\author{Stefano Azzini}%
\affiliation{%
Nanoscience Laboratory, Department of Physics, University of Trento, Italy\\
}%

\author{Sonia Mazzucchi}%
\affiliation{%
 Department of Mathematics and TIFPA-INFN, University of Trento, Italy\\
}%

\author{Valter Moretti}
\affiliation{%
 Department of Mathematics and TIFPA-INFN, University of Trento, Italy\\
}%

\author{Lorenzo Pavesi}%
\affiliation{%
Nanoscience Laboratory, Department of Physics, University of Trento, Italy\\
}%

\maketitle

\section*{Supplementary Note 1: Single photon entanglement, ideal implementation}\label{Section1}

In our experimental implementation (setup shown in Fig. 2a of the main text) we use the momentum and polarization degrees of freedom of a single photon. This system can be represented as a 2-qubit system: the two directions of propagation $\{0,1\}$ are two orthonormal basis vectors in the momentum Hilbert space $\cH_M$ while the two polarizations $\{V,H\}$, vertical and horizontal, correspond to two orthonormal basis vectors in the Hilbert space $\cH_P$. Consequently, the four vectors $\{|0H\rangle, |0V\rangle, |1H\rangle, |1V\rangle\}$ are an orthonormal basis in the tensor product Hilbert space $\cH_M\otimes\cH_P$. The two polarization beamsplitters and the four detectors in the measurement stage (Fig. 2a of the main text) implement the projection-valued measure (PVM) corresponding to the four orthogonal projectors: 
\begin{equation}|0V\rangle\langle 0V|, \quad  |0H\rangle\langle 0H|, \quad  |1V\rangle \langle 1V|, \quad  |1H\rangle\langle 1H|\label{PVM}.\end{equation}
Defined $\{ P^M_{+1}, P^M_{-1}\}$ ( $\{ P^P_{+1}, P^P_{-1}\}$ ) the PVM associated to the momentum observable $O^M=\sigma_z$ (polarization observable $O^P=\sigma_z$), the projectors in \eqref{PVM} can be written in product form as $P^M_a\otimes P^P_b$, where the superscripts M(P) stands for momentum (polarization), and $a, b\in \{+1, -1\}$.
For any unit vectors $\boldsymbol u, \boldsymbol v \in \bR^3$, it is possible to define $U_{\boldsymbol u} :\cH_M\to \cH_M$ and $U_{\boldsymbol v} :\cH_P\to \cH_P$, unitary operators that transform the PVM of the operator $O^M=\sigma _z$ and $O^P=\sigma _z$ into the PVM $\{ P^{\boldsymbol u}_{+1}, P^{\boldsymbol u}_{-1}\}$ and $\{ P^{\boldsymbol v}_{+1}, P^{\boldsymbol v}_{-1}\}$ of the operators $\boldsymbol u\cdot \sigma$ and $\boldsymbol v\cdot \sigma$:
\begin{equation}
P^{\boldsymbol u}_{+1}=U^\dag_{\boldsymbol u} P^M_{+1}U_{\boldsymbol u}, \quad \mathbb{P}^{\boldsymbol u}_{-1}=U^\dag_{\boldsymbol u} P^M_{-1}U_{\boldsymbol u}.
\end{equation}
\begin{equation}
P^{\boldsymbol v}_{+1}=U^\dag_{\boldsymbol v} P^P_{+1}U_{\boldsymbol v}, \quad \mathbb{P}^{\boldsymbol v}_{-1}=U^\dag_{\boldsymbol v} P^P_{-1}U_{\boldsymbol v}.
\end{equation}
Here $\sigma=(\sigma_x,\sigma_y,\sigma_z)$, where $\{\sigma_i\}_{i=x,y,z}$ are the Pauli matrices, and $\boldsymbol u\cdot \sigma:=u_x\sigma_x+u_y\sigma_y+u_z\sigma_z$.
For a generic state $\rho$ in $\cH_M\otimes \cH_P$, the probability of the outcomes $(a,b)$ for the joint measurements of the observable $O^{\boldsymbol u}={\boldsymbol u}\cdot \sigma$ (for the momentum degree of freedom) and of the  observable $O^{\boldsymbol v}={\boldsymbol v}\cdot \sigma$ (for the polarization degree of freedom) is given by 
\begin{eqnarray} \label{det-prob-Sup}
\mathbb{P}(a,b&|&\rho, {\boldsymbol u}, {\boldsymbol v})=   \Tr[\rho P^{\boldsymbol u}_{a}\otimes P^{\boldsymbol v}_{b}]\label{p-1}\\
&=& \Tr[\rho U^\dag_{\boldsymbol u} P^M_{a}U_{\boldsymbol u}\otimes U^\dag_{\boldsymbol v} P^P_{b}U_{\boldsymbol v}]\\
&=& \Tr[\rho ( U^\dag_{\boldsymbol u}\otimes  U^\dag_{\boldsymbol v}) \,( P^M_{a}\otimes P^P_{b})\, ( U_{\boldsymbol u}\otimes  U_{\boldsymbol v})]\\
&=& \Tr[(U_{\boldsymbol u}\otimes  U_{\boldsymbol v}) \rho ( U^\dag_{\boldsymbol u}\otimes  U^\dag_{\boldsymbol v}) \, P^M_{a}\otimes P^P_{b} ] \label{p-4}
\end{eqnarray}
The last line shows that the statistical distribution of the measurement of $\{P^{\boldsymbol u}_{a}\otimes P^{\boldsymbol v}_{b}\}_{a,b=\pm 1}$ on a state $\rho $ coincides with the distribution of outcomes of the measurement described by the PVM \eqref{PVM} over the rotated state $\rho_{\boldsymbol u, \boldsymbol v}$, which is given by:
\begin{equation}
\rho_{{\boldsymbol u}, {\boldsymbol v}}=U_{\boldsymbol u}\otimes  U_{\boldsymbol v} \, \rho\,  U^\dag_{\boldsymbol u}\otimes  U^\dag_{\boldsymbol v} 
\end{equation}

The first part of the preparation stage shown in Fig. 2a (orange box) of the main text, consisting of a Mach Zehnder interferometer (MZI), realizes the rotation operator $U_{\boldsymbol u} \otimes I$, while the second part, consisting in two half-wave plates, implements the operator $I\otimes U_{\boldsymbol v}$. Actually, the first part sets the measurement basis of the momentum observable $O^{\boldsymbol u}={\boldsymbol u}\cdot \sigma$, while the second part  does the same for the polarization observable $O^{\boldsymbol v}=\boldsymbol v\cdot \sigma$.

The rotation operation for the momentum $U_{\boldsymbol u}$, in the case in which lossless balanced beam splitters (BSs) are used, is unitary and can be obtained as $U_{\boldsymbol u}=V_{BS}
V(\phi)V_{Mr}V_{BS}$, where:
\begin{equation*}
\begin{aligned}
&V_{BS}=\left(\begin{array}{ll}
\sqrt{0,5}& i\sqrt{0,5} \\
i\sqrt{0,5}& \sqrt{0,5}
\end{array}\right),\\
&V(\phi)=\left(\begin{array}{ll}
e^{2i\phi}&0 \\
0& 1
\end{array}\right),
\\
&V_{Mr}=\left(\begin{array}{ll}
0&1 \\
1& 0
\end{array}\right).
\end{aligned}
\end{equation*}
The last operator $V_{Mr}$ represents the swapping action of the mirrors between the two BSs, in which an identical optical response for the vertical and horizontal polarization is assumed.
Computing $U_{\boldsymbol u}$, we obtain:
$$U_{\boldsymbol u}= \left(\begin{array}{ll}
\frac{i+ie^{2i\phi}}{2} & \frac{e^{2i\phi}-1}{2}  \\
\frac{1-e^{2i\phi}}{2} &\frac{i+ie^{2i\phi}}{2}
\end{array}\right)=ie^{i\phi}\left(\begin{array}{ll}
\cos(\phi)& \sin(\phi) \\
-\sin(\phi)& \cos(\phi)
\end{array}\right).$$
The last formula shows that the rotation angle is induced by the phase mismatch between the two arms of the MZI.

In the polarization Hilbert space, the rotation operator $U_{\boldsymbol v}$, has a form:
$$U_{\boldsymbol v}=\left(\begin{array}{ll}
\cos(\theta)& \sin(\theta) \\
-\sin(\theta)& \cos(\theta)
\end{array}\right),$$
which is analogous to $U_{\boldsymbol u}$.

\section*{Supplementary Note 2: Single photon entanglement, actual implementation}
In this section, we list the actual parameters for the various optical components of the setup (Fig. 2a of the main text). We list the values of the power reflection and transmission coefficients for vertical and horizontal polarization. These have been measured on the used components at the working wavelength of 543.5 nm. Finally, in the last subsection, we report the dead time value, the after-pulsing probability and the dark counts rate of our single photon avalanche detectors (SPADs).

\subsection*{Generation stage}
\begin{figure}[h!]
\centering
\includegraphics{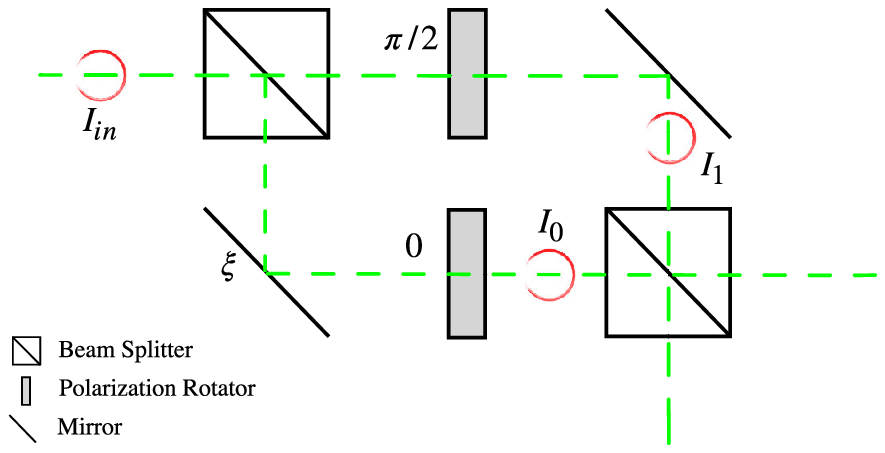}
\caption{\textbf{Details of the generation stage of Fig 2a (main text)}. The points where the optical intensity $I_{in}$,$I_0$ and $I_1$ are acquired are circled in red.}
\label{fig:t0t1}
\end{figure}
The coefficients $t_0$ and $t_1$ of Eq.(13) (main text) are calculated as:
\begin{equation}
    t_0=\sqrt{\frac{I_0}{I_{in}}}, \qquad  t_1=\sqrt{\frac{I_1}{I_{in}}}
\end{equation}
and their values is reported in Table \ref{tab:coeff_gen}.
$I_0$, $I_1$ and $I_{in}$ refer to the optical power measured at the three red circles of Fig.~\ref{fig:t0t1}. These coefficients take into account the transmission of the first beam splitter, the reflection of the mirrors and the transmission of the half-wave plates of the generation stage.

\begin{table}[h!]
\centering
\begin{tabular}{c|c}
\hline
$|t_0|^2$ & $0.421\pm0.002$   \\
$|t_1|^2$ & $0.456\pm0.001$  \\
\hline
\end{tabular}
\caption{\textbf{Measured transmission coefficients relative to the optical paths $|0\rangle$ and $|1\rangle$ in the generation stage.} For each optical path in the generation stage of Fig. 2a (main text), the measured values at 543.5 nm are reported.}
  \label{tab:coeff_gen}
\end{table}

\subsection*{Rotation stage}
In Table~\ref{tab:coeff_bs} we report the measured values of transmission and reflection coefficients of the two beam splitters that compose the Mach Zehnder Interferometer (MZI) in Fig. 2a (main text), for vertical and horizontal polarization.
\begin{table}[h!]
\centering
\begin{tabular}{c|cc}
\hline
&$BS_1$&$BS_2$\\
\hline
$|t_V|^2$ &  $0.502\pm0.005$   & $0.476\pm0.003$ \\
$|r_V|^2$ &  $0.423\pm0.003$   & $0.416\pm0.001$ \\
$|t_H|^2$ &  $0.511\pm0.002$   & $0.4865\pm0.001$ \\
$|r_H|^2$ &  $0.349\pm0.001$   & $0.3583\pm0.0007$ \\
\hline
\end{tabular}
\caption{\textbf{Power transmission and reflection coefficients for the two beam splitters composing the MZI and for the two polarizations.} The measurements were done at 543.5 nm and for two incident light polarizations (V vertical, H horizontal).}
  \label{tab:coeff_bs}
\end{table}

\begin{figure}[h!]
	\centering
	\includegraphics{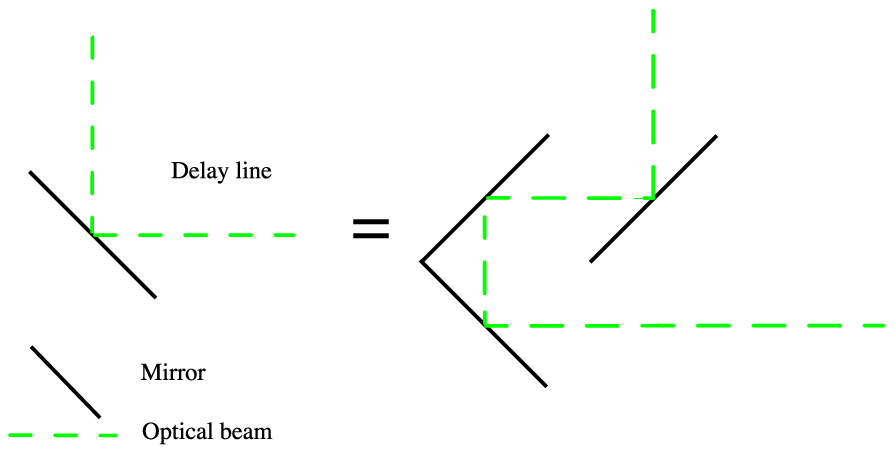}
	\caption{\textbf{Delay Line.} Scheme of the DL mirrors of the MZI of Fig. 2a (main text). The DL mirrors are actually composed by three mirrors as shown here.}
	\label{fig:dl}
\end{figure}

The mirrors in the MZI of Fig. 2a (main text) were composed by a series of mirrors to ease the optical alignment, as reported in Fig.~\ref{fig:dl}. In the figure they were named delay line since they allow to increase/decrease one path length with respect to the other. The measured power transmission coefficients for the two polarizations of these components are reported in Table~\ref{tab:coeff_mirr}.

\begin{table}[h!]
\centering
\begin{tabular}{c|cc}
\hline
&$DL_1$&$DL_2$\\
\hline
$|\gamma_{V}|^2$ & $0.898\pm0.005$ & $0.872\pm0.006$ \\
$|\gamma_{H}|^2$ & $0.798\pm0.004$ & $0.771\pm0.002$ \\
\hline
\end{tabular}
\caption{\textbf{Power transmission coefficients for the two delay lines used in the MZI.} For each delay line in the MZI (DL in Fig. 2 (main text)) we report its measured transmission coefficient for the selected polarization.}
  \label{tab:coeff_mirr}
\end{table}

\subsection*{Detection stage}
In Table~\ref{tab:det} we report the typical values of dead time, afterpulsing probability and dark count rate for our SPAD, as given by the manufacturer.
\begin{table}[h!]
\centering
\begin{tabular}{c|ccc}
\hline
 &$T_d$ [ns]&$p_a$[\%]& DCR [Hz]\\
\hline
SPAD &22&0.5&$<100$ \\
\hline
\end{tabular}
\caption{\textbf{SPAD characteristics.} Typical values of dead time ($T_d$), after-pulsing probability($p_a$) and dark count rate(DCR) for the SPADs used in the experiment.}
  \label{tab:det}
\end{table}

\section*{Supplementary Note 3: Treatment of the non-idealities of the optical components in estimating the guessing probabilities} 
The device independent bound developed in \cite{Pironio10} for the guessing probability relies on the product form of the PVM $\{P^{\boldsymbol u}_{a}\otimes P^{\boldsymbol v}_{b}\}_{a,b=\pm 1}$ describing the joint measurement of the two observables $O^{\boldsymbol u}$ and $O^{\boldsymbol v}$. When considering space-like separated or shielded systems, this condition is fulfilled unless the Bell test is affected by the locality loophole.
However, more realistic and technologically accessible implementations require an additional set of rather mild assumptions. As discussed in the main text, we assume that the manufacturer of the devices is trusted and that the technical features of the components fulfill a few additional requirements. Specifically, all components present some non-idealities and these must be taken into account in an estimation of the effective guessing probability. In \cite{Mazzucchi2021} a theoretical justification of the methodology here described is provided.

For the implemented QRNG based on SPE, the product form of the rotation operator $U_{\boldsymbol u} \otimes U_{\boldsymbol v}$ is essential to guarantee the product form of the PVM $\{P^{\boldsymbol u}_{a}\otimes P^{\boldsymbol v}_{b}\}_{a,b=\pm 1}$. As discussed in Supplementary Note 1, the rotation operation $U_{\boldsymbol u} \otimes U_{\boldsymbol v}$ is obtained using a MZI and two half-wave plates, which independently rotate the momentum and polarization qubits. Respectively, the MZI implements the operation $U_{\boldsymbol u} \otimes I$,  and the two half-wave plate $I \otimes U_{\boldsymbol v}$. While the real description of the two half-wave plates do not present issues regarding the implementation of the product form of $I \otimes U_{\boldsymbol v}$, the real description of the MZI presents some problems regarding the realization of $U_{\boldsymbol u} \otimes I$. In particular the beam splitters (BSs) and the mirrors in the MZI (see Fig. 2a main text) present some polarization non-idealities. Specifically, the transmission and the reflection coefficients for the two polarizations, vertical and horizontal, are different. The matrix representation of the BS in the basis  $\{|0V\rangle,  |0H\rangle, |1V\rangle, |1H\rangle\}$ can be modeled as:

\begin{equation}\label{BSM}
\begin{aligned}
U_{BS}^{real}=&\left(\begin{array}{llll}
t_V& 0 & ir_V &0\\
0 & t_H & 0 & ir_H\\
ir_V & 0 & t_V & 0\\
0 & ir_H & 0 & t_H
\end{array}\right), \\
&\begin{array}{l}
|t_V|^2+|r_V|^2\leq 1\\
|t_H|^2+|r_H|^2\leq 1\\
\end{array} 
\end{aligned}
\end{equation}
where $t_{V,H},r_{V,H}$ are the transmission and reflection coefficients of the BSs for the corresponding polarization.
Analogously, the matrix representing the action of the mirrors can be written as:
\begin{equation}\label{MrinesM}
\begin{aligned}
U_{Mr}^{real}=&\left(\begin{array}{llll}
0 & 0 & \gamma_{V_1} & 0\\
0 & 0 & 0 & \gamma_{H_1}\\
\gamma_{V_2} & 0 & 0 & 0\\
0 & \gamma_{H_2} & 0 & 0
\end{array}\right),\\
&\begin{array}{l}
|\gamma_{x_y}  |^2\leq 1,\\
\forall x\in\{V,H\},\forall y\in\{1,2\}
\end{array} 
\end{aligned}
\end{equation}
where $\gamma_{x_y}$ is the transmission coefficient for the polarization $x$ of the mirror y of the MZI. Due to these non-idealities, the rotation operator $U^{real}_{{ \phi}, {\theta}}$\footnote{To simplify the notation, we replace $\boldsymbol u$ with the momentum rotation angle $\phi$, and $\boldsymbol v$ with the polarization rotation angle $\theta$.}, is no more the tensor product $U_{\phi}\otimes U_{\theta}$ as in the ideal case. 
Moreover, the rotation operator $U^{real}_{{\phi}, {\theta}}$ is not unitary: due to the presence of losses ( absorption and scattering for the different optical components ) some of the photons are actually lost during the propagation into the experimental setup. Consequently, the statistics of detection outcomes is determinated only by the revealed photons and the probabilities of the possible outcomes $(a,b)$ have to be estimated as:
\begin{equation}\label{p-real}
\mathbb{P}^{real}(a,b|\rho,\phi,\theta)=\frac{ \Tr[U^{real}_{{\phi}, {\theta}}\rho (U^{real}_{{\phi}, {\theta}})^\dag \, P^M_{a}\otimes P^P_{b}]  }{\Tr[U^{real}_{{\phi}, {\theta}}\rho (U^{real}_{{\phi}, {\theta}})^\dag ] },
\end{equation}
where 
\begin{equation}\label{U-real-1}U^{real}_{\phi,\theta}=(I_2\otimes U_{\theta})U_{BS_2}^{real}(V(\phi)\otimes I_2)U_{Mr}^{real}U_{BS_1}^{real} .\end{equation}

In the last equation we have considered that we have two different BSs with different values of $t_V, t_H, r_V, r_H$:
\begin{equation}
\begin{aligned}
&U_{BS_1}^{real}=\left(\begin{array}{llll}
t_{V_1}& 0 & ir_{V_1} &0\\
0 & t_{H_1} & 0 & ir_{H_1}\\
ir_{V_1} & 0 & t_{V_1} & 0\\
0 & ir_{H_1} & 0 & t_{H_1}
\end{array}\right), \\
&U_{BS_2}^{real}=\left(\begin{array}{llll}
t_{V_2}& 0 & ir_{V_2} &0\\
0 & t_{H_2} & 0 & ir_{H_2}\\
ir_{V_2} & 0 & t_{V_2} & 0\\
0 & ir_{H_2} & 0 & t_{H_2}
\end{array}\right).
\end{aligned}
\end{equation}

Due to all these non-idealities,
the probabilities $\mathbb{P}^{real}(a,b|\rho,\phi,\theta)$ differ from  \eqref{p-4} or  \eqref{p-1}. 

We recall that the uniform bound developed in \cite{Pironio10} refers to the maximum attainable (realization independent) guessing probability for the outcomes of measurements of observables of the product form. To cope with this fact, we numerically compute a bound, $e_P$, for the difference between the real probabilities \eqref{p-real} and the probabilities:
\begin{equation}
\mathbb{P}^{ideal}(a,b|\rho,\phi,\theta)=\Tr[\tilde U^{ideal}_{\phi,\theta}\rho (\tilde U^{ideal}_{\phi,\theta})^\dag \, P^M_{a}\otimes P^P_{b}] ,
\end{equation}with : 
\begin{equation}\label{utildereal-bis}\tilde U^{ideal}_{\phi,\theta}= (I\otimes U_{\theta})\tilde U_{BS_2}^{ideal}(V(\phi)\otimes I_2 ) (V_{Mr}\otimes I_2)\tilde U_{BS_1}^{ideal}  .\end{equation}
The unitary operators $\tilde U_{BS_1}^{ideal}$ and $\tilde U_{BS_2}^{ideal}$ are defined as
\begin{equation}
\begin{aligned}
&\tilde U_{BS_1}^{ideal}=\left(\begin{array}{llll}
\cos{\alpha}& 0 & i\sin{\alpha} &0\\
0 & \cos{\alpha} & 0 & i\sin{\alpha}\\
i\sin{\alpha} & 0 & \cos{\alpha} & 0\\
0 & i\sin{\alpha} & 0 & \cos{\alpha}
\end{array}\right) ,\\
&\tilde U_{BS_2}^{ideal}=\left(\begin{array}{llll}
\cos{\beta}& 0 & i\sin{\beta} &0\\
0 & \cos{\beta} & 0 & i\sin{\beta}\\
i\sin{\beta} & 0 & \cos{\beta} & 0\\
0 & i\sin{\beta} & 0 & \cos{\beta}
\end{array}\right)
\end{aligned}
\end{equation}   
Please note that, for each choice of $\alpha,\beta \in [0,\frac{\pi}{2}]$, the two operators $U_{BS_{1,2}}^{ideal}$ are in the product form, which means that the operator $\tilde U^{ideal}_{\phi,\theta}$ is in product form. The probabilities $\mathbb{P}^{ideal}(a,b|\rho,\phi,\theta)$ correspond, indeed, to observables in product form.
This fact enters directly in the calculation of $e_P$,  defined as:
\begin{equation}
\begin{aligned}
&e_P=\\
&\min_{\{\alpha,\beta\}}\left[\max_{\{\phi,\theta,\rho,a,b\}}\left|\frac{ \Tr[U^{real}_{\phi,\theta}\rho (U^{real}_{\phi,\theta})^\dag \, P^M_{a}\otimes P^P_{b}]  }{\Tr[U^{real}_{\phi,\theta}\rho (U^{real}_{\phi,\theta})^\dag ] }-\right.\right.\\
&\left.\left.\Tr[\tilde U^{ideal}_{\phi,\theta}\rho (\tilde U^{ideal}_{\phi,\theta})^\dag \, P^M_{a}\otimes P^P_{b}] \right|\right].
\end{aligned}
\end{equation}
Indeed, since the bound of \cite{Pironio10} is realization independent, we can choose the best couple of $(\alpha,\beta)$ that minimizes the maximum distance over the parameters $\{\phi,\theta,\rho,a,b\}$.

The same reasoning can be applied to the correlation function $I$. Indeed we can define:
\begin{equation}
\begin{aligned}
    &I^{ideal}(\phi_0,\phi_1,\theta_0,\theta_1)=\\
    &\sum_{x,y}(-1)^{xy}\left(\mathbb{P}^{ideal}(a=b|\rho,\phi_x,\theta_y)-\mathbb{P}^{ideal}(a\neq b|\rho,\phi_x,\theta_y)\right).
    \end{aligned}
\end{equation}
and
\begin{equation}
\begin{aligned}
    &I^{real}(\phi_0,\phi_1,\theta_0,\theta_1)=\\
    &\sum_{x,y}(-1)^{xy}(\mathbb{P}^{real}(a=b|\rho,\phi_x,\theta_y)-\mathbb{P}^{real}(a\neq b|\rho,\phi_x,\theta_y)).
    \end{aligned}
\end{equation}
As before, we define 
\begin{equation}
\begin{aligned}
&e_I=\\
&\min_{\{\alpha,\beta\}}\left[\max_{\{\phi_0,\phi_1,\theta_0,\theta_1,\rho,a,b\}}\left| I^{ideal}(\phi_0,\phi_1,\theta_0,\theta_1)-\right.\right.\\
&\left.\left.I^{real}(\phi_0,\phi_1,\theta_0,\theta_1) \right|\right],
\end{aligned}
\end{equation}
where we take the minimum value over $(\alpha,\beta)$ for the maximum over the parameters $\{\phi_0,\phi_1,\theta_0,\theta_1,$$\rho,a,b\}$.

Lastly we need to model the state $\rho$, Eq. (10) main text:
\begin{equation}\label{eq:Model_Rho}
\rho(v,\delta,\pi_1,\pi_2)=R(\pi_1,\pi_2)\rho_s(v,\delta)R(\pi_1,\pi_2)^{\dagger}.
\end{equation}
Here, we give the explicit description of $R(\pi_1,\pi_2)$:
\begin{equation}
R(\pi_1,\pi_2)=\left(\begin{array}{llll}
\cos(\pi_1)& \sin(\pi_1) & 0 &0\\
-\sin(\pi_1) & \cos(\pi_1) & 0 & 0\\
0 & 0 & \cos(\pi_2)& \sin(\pi_2)\\
0 & 0 & -\sin(\pi_2)& \cos(\pi_2)
\end{array}\right).
\end{equation}
In addition, we write also $\rho_s(v,\delta)$ in the matrix form:
\begin{equation}
\begin{aligned}
&\rho_s(v,\delta)=\\
&\left(\begin{array}{llll}
v|t_{0n}|^2+(1-v)& 0 & 0 &vt_{0n}(t_{1n}e^{i\delta})^*\\
0 & 1-v & 0 & 0\\
0 & 0 & 1-v& 0\\
vt_{0n}^*t_{1n}e^{i\delta} & 0 & 0& v|t_{1n}|^2+(1-v)
\end{array}\right),
\end{aligned}
\end{equation}
where we recall that $t_{0n}$ and $t_{1n}$ are the normalized transmission coefficients of the two paths $|0\rangle$ and $|1\rangle$, as reported in Eq. (12) (main text).
In the most general scenario, $\rho$ can be written as:
\begin{equation}
    \rho=(A+iB)^{\dagger}(A+iB)
\end{equation}
where A and B are
\begin{equation} 
\begin{aligned}
&A=\left(\begin{array}{llll}
x_1& x_5 & x_8 &x_{10}\\
0 &x_2 & x_6 & x_9\\
0 & 0 & x_3 & x_7\\
0 & 0 & 0 & x_4
\end{array}\right) ,\\
&B=\left(\begin{array}{llll}
0& x_{11} & x_{14} &x_{16}\\
0 & 0 & x_{12} & x_{15}\\
0 & 0 & 0 & x_{13}\\
0 & 0 & 0 & 0
\end{array}\right).
\end{aligned}
\end{equation}
To ensure that $\rho$ is semi-definite positive and normalized, we have used the Cholesky decomposition~\cite{Golub96} and the spherical 16-dimensional coordinates $\{x_i\}_{i=1..16}$.
The latter are unequivocally fixed by the choice of the radius r, which is fixed to the value 1 and by 15 angles $\{\eta_i\}_{i=1..15}$:
\begin{equation}
\begin{split}
    x_1&=\cos(\eta_1),\\
    x_2&=\cos(\eta_2)\sin(\eta_1),\\
    x_3&=\cos(\eta_3)\sin(\eta_2)\sin(\eta_1),\\   
    x_4&=\cos(\eta_4)\sin(\eta_3)\sin(\eta_2)\sin(\eta_1),\\
    x_5&=\cos(\eta_5)\sin(\eta_4)...\sin(\eta_1),\\
    ...\\
    x_{15}&=\cos(\eta_{15})\sin(\eta_{14})...\sin(\eta_1),\\
    x_{16}&=\sin(\eta_{15})...\sin(\eta_1).\\
\end{split}
\end{equation}
To ensures that the diagonal terms $x_1,x_2,x_3$ and $x_4$ are positive, we restrict the angles $\{\eta_i\}_{i=1..4}$ to $[0,\pi/2]$. The other angle domains are $\{\eta_i\}_{i=5..14}\in [0,\pi]$ and $\eta_{15}\in [0,2\pi]$.

\subsection*{Numerical evaluation}
The maximization procedure over the parameters $\{\phi_0,\phi_1,\theta_0,\theta_1,$ $\rho,a,b\}$ for $e_I$ and $\{\phi,\theta,\rho,x,y\}$ for $e_P$ has been performed using the Sequential Quadratic Programming (SQP). This was implemented using as Global Optimization Toolbox the MultiStart solver by Matlab(c), using the standard SQP algorithm.
The operators $U_{BS_1}^{real}$,$U_{BS_2}^{real}$,$U_{Mr}^{real}$ are built using the measured values of transmission and reflection coefficients for our experimental setup.
Since the problem necessitates to minimize the upper bounds for ($\alpha,\beta$) we have followed this procedure:
\begin{itemize}
    \item Fix a pair of ($\alpha,\beta$)
    \item Perform a rough maximization over the variable parameters, calculating $e_I$ and $e_P$
    \item Calculate the guessing probability
    \item Repeat
\end{itemize}
The pair of angles ($\alpha,\beta$) which gives the lowest guessing probability is the best one. To ensure that the result we have found is accurate, we have repeated the optimization starting from several initial points: the maximum value obtained was considered to be the global maximum of the function. We have decided to set the number of starting points to $3\times 10^3$ for the optimization of $e_P$ and $10^4$ for $e_I$. The Matlab solver converges to the values reported in the main text. The errors on $e_P$ and $e_I$ are estimated in the this way: for each experimental quantity measured ( transmission and reflection coefficients of the optical components ) we create a normal distribution having as mean the measured value, and as standard deviation the error of the measurements. From each of these distributions we randomly extract one value and we calculate $e_P$ and $e_I$ keeping fixed the values of the variable parameters that maximize $e_I$ and $e_P$ obtained before. In this way, we can find an error which depends only on the experimental errors over the measurement performed on the optical components. This operation is repeated $4 \times 10^3$ times and we take as error the standard deviation of these obtained values. We must remark that the important part of this numerical evaluation is the maximization part, which is necessary to guarantee a bound which is independent of the parameters, while the minimization impacts only on the efficiency of the QRNG and not on its security.

\section*{Supplementary Note 4: Markovian Model for detectors non-idealities}

Here, we introduce the Markovian model used to estimate the detection probabilities $\mathbb{P}^{real}(a,b|\rho,\phi,\theta)$ (see \eqref{p-real}) from the experimental data. 

Set a particular choice of the pair of parameters $\phi, \theta$, we denote the four probabilities $\{\mathbb{P}^{real}(a,b|\rho,\phi,\theta)\}_{a,b=\pm 1}$ with the symbols $p_j$, $j=1,2,3,4$, to ease the following discussion. Besides the notational simplicity, this choice is motivated by the consideration that each of the four values provides the probability that the incoming  photon is detected by the $j_{th} $ detector present in the final stage of the experimental setting.

In a realistic implementation, as discussed in the main text, the presence of afterpulsing  and detector dead time produces correlations among the subsequent readouts of the detectors, which in fact can no longer be consider independent. These memory effects can be described in terms of a  Markovian model, which is based on the following set of  assumptions:
\begin{itemize}
\item The source has a poissonian emission statistics with effective intensity parameter $\lambda_{e }=\eta \lambda$, where $\eta<1$ is the detector efficiency, assumed equal for the four SPADs.
\item $p_a$, the probability of after-pulsing, is 0.5 \% or less.
\item The dead time of detectors, $T_d$, is negligible with respect to the average inter-arrival time $1/\lae$. More precisely, we assume that the product $\lae T_d$ is of the order $10^{-2}$ or less.
\item The time of after-pulsing $T_a$ and the dead time $T_d$ are of the same order ($T_a \sim T_d$).
\end{itemize}

In the following the sequence of subsequent readouts of the detectors  will be  described in terms of a sequence  $\{\eta_n\}_n$   of random variables with four possible outcomes $i=1,2,3,4$  (i.e. $\eta_n=i$ means that the $n_{th}$ incoming photon was detected by the $i_{th}$ SPAD).   
According to the approximations above, it is possible to prove (see \cite{Mazzucchi2021} for further details) that the sequence $\{\eta_n\}$ is  a stationary Markov chain, i.e. the probability of predicting the $n_{th} $ readout given the previous $(n-1)$ is equal to
\begin{align*}
&\bP(\eta_{n}={i_n}|\eta_{n-1}=i_{n-1},\ldots \eta_{0}=i_{0 } )\\ &=\bP(\eta_{n}=i_n|\eta_{n-1}=i_{n-1})\\
&=\bP(\eta_{1}=i_n|\eta_{0}=i_{n-1}), \qquad \forall i_0, \ldots i_{n-1}, i_n=1,2,3,4 .
\end{align*}
Further, the  transition probabilities $\bP(\eta_{n+1}=j|\eta_{n}=i )=P_{ij}$ are given by:
\begin{equation}\label{Pij}
P_{ij}=p_a\delta_{ij}+(1-p_a )\left((1-\lae T_d) p_j+\lae T_d q_{ij}\right)\, , 
\end{equation}
where $q_{ij}=p_i^2\delta_{ij}+(1+p_i)p_j(1-\delta_{ij})$, $\delta$ being the Kronecker delta.
An unbiased estimator for the four theoretical probabilities $p_i$, $i=1,2,3,4$ can be constructed via the maximum likelihood principle. 
Given a realization of the Markov chain described above, i.e. a sequence of outcomes $\{x_i\}_{i=1,\ldots, n}$, with $x_i=1,2,3,4$, its probability is given by 
$p_{x_1}\prod_{i=1}^{n-1}P_{x_ix_{i+1}}$
and the corresponding log-likelihood  is equal to
$$l(P):=\log(p_{x_1}\prod_{i=1}^{n-1}P_{x_ix_{i+1}})=\log (p_i)+\sum_{i,j=1,2,3,4}N_{ij}\log(P_{ij}),$$
where $N_{ij}$ is the number of transitions from $i$ to $j$. The function $l(P)$ has to be maximized under the four constraints $\sum_jP_{ij}=1$, $i=1,2,3,4$, providing unbiased estimators  $\hat p_i$ for the probabilities $p_i$, $i=1,2,3,4$. The corresponding confidence intervals are evaluated according to the the technieques presented in \cite{billingsley1961statistical}.

In fact, the estimated values $\hat p_i$ as well as the corresponding confidence intervals were computed using a code written using the statistical programming language R.

The Markovian model plays an important role also in the estimation of the guessing probability present in the sequence of raw data. Indeed, the semi-device independent entropy certification protocol described in the main paper yields a uniform bound for the theoretical detection probabilities \eqref{p-real}. In particular, 
setting $p_{max}:=\frac{1}{2}+\frac{1}{2}\sqrt{2-(|\hat{I}_{real}|-e_I)^2/4} + e_P$ we have $p_j\leq p_{max}$
for any choice of the pair $\phi,\theta$ and for any detection channel $j=1,\ldots, 4$.

On the other hand, taking into account the non-idealities of the detectors and the resultant correlations between subsequent photon detections, the effective guessing probability present in the sequence of the raw data has to be estimated as the maximum value of the probability of the $n_{th}$-readout given the previous ones:
\begin{equation}
	p^*_{guess}:=\sup_{i_0,\ldots ,i_n}\bP(\eta_{n}={i_n}|\eta_{n-1}=i_{n-1},\ldots \eta_{0}=i_{0 })  .
\end{equation}

By the Markov property, this reduces to
\begin{equation*}
\begin{aligned}
&p^*_{guess}=\sup_{i,j}P_{ij}=\\
&\sup _{i,j}\left(p_a\delta_{ij}+(1-p_a )\left((1-\lae T_d) p_j+\lae T_d q_{ij}\right)\right).
\end{aligned}
\end{equation*}
By using the inequality $\sup _j p_j\leq p_{max}$ we finally get $p^*_{guess}\leq M(p_{max})$, where the function $M$ is defined as:
\begin{align*}
    &M(p_{max}):=\\
    &\sup_{\begin{subarray}{l}i,j=1,\ldots , 4\\
    p_j\leq p_{max}\end{subarray}}\left(p_a\delta_{ij}+(1-p_a )\left((1-\lae T_d) p_j+\lae T_d q_{ij}\right)\right)\\
    & =\max \left\{    p_a+(1-p_a )\left((1-\lae T_d) p_{max}+\right.\right.\\
    &\left.\left.\lae T_d p_{max}^2\right), (1-p_a)(p_{max}+\lae T_dp_{max}(1-p_{max}))\right\}.
\end{align*}

\bibliography{Reference-SM}